# Controlled fabrication of single electron transistors from single-walled carbon nanotubes

Paul Stokes and Saiful I. Khondaker[1]

Nanoscience Technology Center & Department of Physics, University of Central Florida, 12424 Research Parkway, Orlando FL 32826, USA

**Abstract**

Single electron transistors (SETs) are fabricated by placing single walled carbon nanotubes (SWNTs) on a 100 nm wide local Al/Al$_2$O$_3$ bottom gate and then contacting with Pd electrodes. Coulomb oscillations up to 125 K were observed and charging energies of 12-15 meV with level spacing of ~ 5 meV were measured from the Coulomb diamond, in agreement with a dot size of ~100 nm, implying that the local gate defines the dot size by bending SWNT at the edges and controls its operation. This "mechanical template" approach may facilitate large scale fabrication of SET devices using SWNT.

Single electron transistors (SETs) have attracted considerable attention because of their potential as a building block for future quantum based nanoelectronic devices.[1, 2] A SET consists of a small conducting island connected to two metallic leads through tunnel barriers. Electron tunneling can be controlled one by one with a nearby gate electrode, capacitively coupled to the island when its charging energy ($e^2/C$) is greater than the thermal energy ($k_BT$). By reducing the size of the island, the capacitance decreases leading to a higher charging energy and operating temperature. Since the first SET was demonstrated about 20 years ago in an aluminum tunnel junction[3], it has been realized in a variety of systems including lithographically defined dots in GaAs/AlGaAs heterojunction, direct etching of Si substrate, metallic grains in nanopore, and colloidal nanocrystals.[4-7] Lithography defined dots are limited by the resolution of lithography and often are larger in size, requiring sub Kelvin temperature for operation. In addition, complex fabrication processes makes it difficult to control the uniformity and reproducibility. Metallic grains and colloidal nanocrystals give smaller sized uniform dots with SET operating temperature ~100 K, however, for SET operation they need to be placed in nanosized gaps, which is highly challenging and difficult to control giving an extremely low device yield. Recently, nanowires[8-9] and single walled carbon nanotubes (SWNTs)[10-16] have been considered to be good candidates for the fabrication of SETs because of their small diameters. Fabrication of SET using SWNT relies on the introduction of tunnel barriers. It has been shown that when a SWNT is bent at a selected position, the bend acts as a nanometer sized tunnel barrier.[12-14] By creating a pair of bends on an individual SWNT using AFM tip, SETs have been demonstrated.[12,13] However, AFM manipulation is time consuming and reproducibility of the same sized device can be challenging.

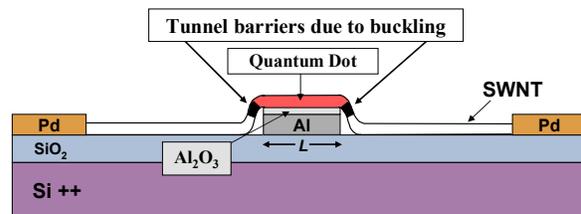

**Figure 1:** (Color online) (a) Schematic diagram of the nanotube SET device. The nanotube bends at the edges of the Al/Al$_2$O$_3$ gate electrode to create two tunnel barriers (black) a distance ($L$) apart. The central island (red) in between the tunnel barriers above the aluminum oxide, is the defined quantum dot. The gate defines the quantum dot and controls its operation.

---

[1] Electronic mail: saiful@mail.ucf.edu

Here, we report a simple device engineering approach to fabricate controllable SETs using SWNT. The approach is based on the formation of two tunnel barriers of controllable separation by naturally bending SWNT at the edges of a raised local gate. A cartoon of our device is shown in Figure 1. A SWNT is placed on a 100 nm wide local $Al/Al_2O_3$ bottom gate and then contacted with Pd source and drain electrodes of 1 um separation on $Si/SiO_2$ substrates. The aluminum gate serves three purposes: (i) it acts as a "mechanical template" to define two tunnel barriers at the edges by naturally bending the nanotube due to van der Walls interactions with the substrate[17], (ii) the width of the gate defines the size ($L$) of the quantum dot, and (iii) it acts as a local bottom gate to control the operation of the SET device. Low temperature electronic transport measurements show Coulomb oscillations up to 125 K. The stability diagram shows charging energies of 12-15 meV and energy level spacing of ~ 5 meV. These energies are in agreement with a quantum dot size of ~100 nm, thus verifying the dot is defined and controlled by the local gate.

Our devices are fabricated on heavily doped Si wafers with a 250 nm $SiO_2$ capped layer. Larger features such as contact pads are first defined by photolithography using double layer resist (LOR 3A/Shipley 1813), developed in CD26, followed by thermal evaporation of 5 nm thick Cr and 40 nm Au, and standard lift-off. We then define arrays of 40 μm long and 100 nm wide patters by means of electron beam lithography (EBL) for the Al gates along with reference markers (to later connect the NTs), followed by 40 nm of thermal Al deposition and lift-off in acetone. A thin aluminum oxide layer is created by oxygen plasma treatment for 10 minutes to serve as a gate dielectric. CVD grown SWNTs (cheaptubes.com) are then ultrasonically dispersed in 1,2-dichloroethane (DCE) for ~5 minutes. The average length of the nanotubes after dispersion is 2-5 μm, determined by AFM. The nanotubes are then spun (~1000 rpm) on the substrate containing the array of Al gates. By tapping mode atomic force microscopy (AFM), we locate the nanotubes that pass over the $Al/Al_2O_3$ gate and record their coordinates with respect to the reference markers. Another step of EBL is then implemented to define source (S) and drain (D) top contacts, followed by 25 nm of Pd deposition and lift-off. Pd was used to make good contact with SWNT to avoid additional tunnel barrier formation at the nanotube source-drain interface.[18] Devices are bonded and loaded into a 4 K cryostat for electronic transport measurements. A total of nine devices were measured all with 100 nm local gate.

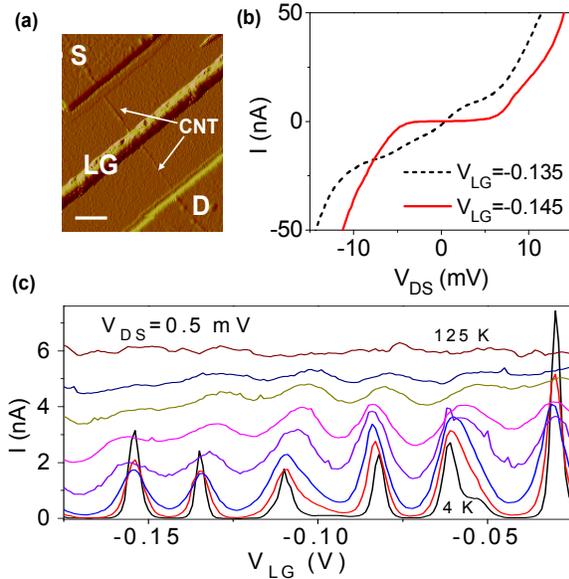

**Figure 2:** (Color online) Device A: (a) Tapping mode atomic force micrograph with 100 nm wide Al local gate (LG) and Pd source (S) and drain (D) top contacts (scale bar is 200 nm). (b) $I$-$V_{DS}$ curves at two different gate voltages at 4.2 K, showing Coulomb Blockade and single electron tunneling. (c) Current versus local gate voltage showing several reproducible peaks for several temperatures (bottom to top: 4.2, 12, 20, 35, 50, 75, 100, and 125 K). Peaks begin to wash out around 125 K, for which $k_BT \sim 11$ meV.

Figure 2a shows an AFM image for one of our devices (Device A). The diameter of the nanotube is ~2 nm determined by AFM height measurement. Room temperature measurements for this device reveal a contact resistance of ~70 kΩ and a small change in current as a function of local gate voltage ($V_{LG}$) indicating a small bandgap nanotube. Gate leakage current is negligible (<1 pA) for a voltage of -1 V to +1 V. Figure 2b is a plot of drain current ($I$) as a function of source drain voltage ($V_{DS}$) for device A measured at 4.2 K for two different local gate voltages $V_{LG}$ = -0.145 V and $V_{LG}$ = -0.135 V. At $V_{LG}$ = -0.145 V, the current is zero between a $V_{DS}$ of -5 mV to +5 mV indicating Coulomb Blockade (CB) behavior. The CB can be lifted by applying $V_{LG}$= -0.135 V. Figure 2c is a plot of current versus $V_{LG}$ for a fixed source drain voltage of $V_{DS}$ = 0.5 mV at various temperatures from 4.2 K up to 125 K for the same device. At 4.2 K the current shows quasi periodic oscillations as a function of $V_{LG}$. It can be seen that as the temperature is raised the peaks start to broaden and ultimately wash out at around 125 K. From this we can estimate the charging energy $U_C$ to be ~11 meV as Coulomb oscillations typically wash out at $T=U_C/k_B$.

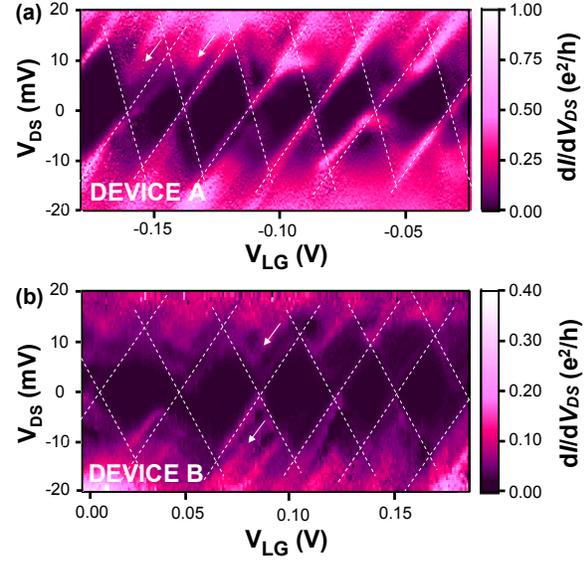

**Figure 3:** (Color online) Differential conductance (d$I$/d$V_{DS}$) as a function of local gate voltage ($V_{LG}$) and source drain voltage ($V_{DS}$) at 4.2 K for two devices (a) device A and (b) device B. Coulomb diamonds are outlined by the white dotted lines for a guide to the eye. The measured charging energy from the diamond height is $U_c$ ~ 13 meV and energy level spacings $\Delta E$ ~ 5 meV are indicated by arrows in both diagrams.

Figure 3a is a plot of differential conductance (d$I$/d$V_{DS}$) for device A, calculated by differentiating the $I$-$V_{DS}$ curves for different gate voltages (0.5 mV step), as a function of both $V_{DS}$ and $V_{LG}$ taken at 4.2 K. The gate voltage range is the same as that is presented in figure 2c. Brighter regions symbolize high conductance (up to ~1.0 e²/h) and darker regions signify Coulomb blockade. Several dark diamond shaped regimes (Coulomb diamonds) are outlined by white dotted lines as a guide to the eye, which are signature of SET. Coulomb diamonds are approximately equally spaced with $\Delta V_G$ ~ 25 mV. Diamonds closing and constant slopes throughout the plots indicate measurement of a single quantum dot.[15] The height of the diamond is a measure of charging energy $U_C$ and additional lines parallel to the boundaries (indicated by arrows) of the Coulomb diamond correspond to single particle level spacing $\Delta E$. From this figure, we measure $U_C$ and $\Delta E$ to be ~13 meV and ~5 meV respectively. Figure 3b shows differential conductance plot of another device (device B) which show similar energy scales.

Using the constant interaction model[19], we can extrapolate several parameters from figure 3. The gate capacitance for Device A can be calculated from $C_G = e/\Delta V_G$ ~ 6.4 aF. We estimate the left and right capacitances from the slopes of the Coulomb diamond, $\alpha_1 = -C_G/C_1$ = -2.2 and $\alpha_2 = C_G/(C_G + C_2)$ = 0.7 yielding $C_1$, $C_2$ ~ 2.9 aF and 2.8 aF respectively. The charging energy $U_C = e^2/C_\Sigma$ can be calculated from the total capacitance of the quantum dot $C_\Sigma = C_1 + C_2 + C_G$. $C_\Sigma = 12.1$ aF, yielding a charging energy $U_C \sim 13.2$ meV, consistent with the value directly read off of the stability diagram from the diamond height and in close agreement with the temperature dependent data. The same calculation for Device B yields $C_\Sigma = 11.9$ aF and $U_C = 13.4$ meV. From the charging energy and energy level spacing, we can estimate the size (*L*) of the dot. For SWNT the charging energy and energy level spacing is expected to be $U_C \sim 1.4 eV/L(nm)$ and *ΔE ~ 0.5 eV/ L(nm)*.[11] From here, for we obtain *L* = 106 nm from the charging energy, *L* = 100 nm from the level spacing, and *L* = 127 nm from the temperature dependence data for Device A. Device B yields *L* = 104 nm from the charging energy and *L* = 100 nm from the level spacing. Three other devices also showed feature of single quantum dot with $U_C$ ranging from 12.2 to 15.0 meV in close agreement with a 100 nm sized dot. This is consistent with the width (100 nm) of our Al gate electrode, thus verifying that the dot is defined and controlled by the local gate. The small variation of charging energy could be due to the small variation in nanotube diameters.

Is it possible that the measured dots are not the "engineered dot" but are accidentally formed due to the random defects? In order to further verify location of the quantum dot, we compare our measured gate capacitances $C_G$ with the geometrical gate capacitance for the cylinder-on-plane configuration *$C_G=2\pi\varepsilon_{avg}\varepsilon_0 L/cosh^{-1}(1+t/r)$*, where $\varepsilon_{avg}$ is the average dielectric constant between air and $Al_2O_3$, *t* is the oxide thickness and *r* is the radius of the tube. With *t* = 5 nm and *r* = 0.75 - 1.5 nm, we obtain values of $C_G$ ranging from ~ 6.1 to 7.7 aF in reasonable agreement with our measured values of 2.0 to 6.5 aF. The close agreement of the measured energy scale with the defined ~100 nm sized dots, along with the agreement between the measured gate capacitances and the geometrical gate capacitance in several samples strongly indicate that quantum dot is indeed defined using our mechanical template approach. It is nevertheless possible that accidental dots can also occur alongside the engineered dot giving rise to multi-dot features[20] and four other samples we measured show such multi-dot features.

In conclusion, we presented a simple device engineering approach for the controlled fabrication of SET using SWNT by employing $Al/Al_2O_3$ local gate as a mechanical template. In this proof of concept experiment, a SWNT is placed on a 100 nm wide local bottom gate and then contacted with Pd source and drain electrodes. Low temperature electronic transport measurements for several devices show the charging energies and single particle energy level spacing to be 12-15 meV and ~5 meV respectively consistent with a dot size of ~100 nm. This confirm that the local gate electrode (i) acts as a "mechanical template" to bend the nanotube at its edges to introduce tunnel barriers, (ii) its width defines the size of the quantum dot, and (iii) it controls the operation of the SET device. Further scaling of the gate widths may allow room temperature operation and work is in progress to that end. Our device engineering approach, if combined with directed assembly of CNT, may offer large scale fabrication of controllable and scalable CNT-SET devices.

We acknowledge useful discussion with Liwei Liu and Masa Ishigami. This work is supported in part by NSF-CARRER award ECS-0748091.